\documentclass[aps,twocolumn]{revtex4}
\usepackage{graphicx}

\begin{document}

\title{Test of nuclear level density inputs for Hauser-Feshbach model
  calculations}
\author{$^1$A.V.~Voinov\footnote{Electronic address: voinov@ohio.edu},
$^1$S.M.~Grimes, $^1$C.R.~Brune, $^1$M.J.~Hornish, $^1$T.N.~Massey,
$^{1,2}$A.~Salas}

\address{$^1$ Department of Physics and Astronomy, Ohio University,
  Athens, OH 45701, USA}
\address{$^2$ Los-Alamos National Laboratory, P-25 MS H846, Los Alamos,
New Mexico 87545, USA}

\begin{abstract}
The energy spectra of neutrons, protons, and $\alpha$-particles have been
measured from the d+$^{59}$Co and
$^3$He+$^{58}$Fe reactions leading to the same compound nucleus, ${}^{61}$Ni.
The experimental cross sections have been
compared to Hauser-Feshbach model calculations using different input level
density models. None of them have been found to agree with experiment.
It manifests the serious problem with available level density
parameterizations especially those based on neutron resonance spacings
and density of discrete levels. New level densities and corresponding
Fermi-gas parameters have been obtained for reaction product nuclei such as
$^{60}$Ni,$^{60}$Co, and $^{57}$Fe.
\end{abstract}

\maketitle

\section{Introduction}

The nuclear level density (NLD) is an important input for the calculation of reaction cross sections in the framework
of Hauser-Feshbach (HF) theory of compound nuclear reactions. Compound reaction cross sections are needed in many
applications including astrophysics and nuclear data for science and technology. In astrophysics the knowledge of
reaction rates is crucial for understanding nucleosynthesis and energy generation in stars and stellar explosions.
In many astrophysical scenarios, e.g. the r-process, the cross sections required to compute the reaction rates are
in the regime where the statistical approach is appropriate~\cite{rauscher}. In these cases HF calculations are an
essential tool for determining reaction rates, particularly for reactions involving radioactive nuclei which are
presently inaccessible to experiment. HF calculations are likewise very important for other applications, e.g.,
the advanced reactor fuel cycle program~\cite{AFC}.

The statistical approach utilized in HF theory~\cite{houser} requires knowledge of the two quantities for
participating species (see details below). These are the transmission coefficients of incoming and outgoing
particles and level densities of residual nuclei. Transmission coefficients can be obtained from optical model
potentials established on the basis of experimental data of elastic and total cross sections. Because of
experimental constraints, the difference between various sources of transmission coefficients
usually does not exceed $10-15\%$.
Level densities are more uncertain. The reason is that it is difficult to obtain them experimentally above the
region of well-resolved discrete low-lying levels known from nuclear spectroscopy. At present, the level density
for practical applications is calculated mainly on the basis of the Fermi-gas \cite{Bethe} and Gilbert-Cameron
\cite{GC} formulas with adjustable parameters which are found from experimental data on neutron resonance spacing
and the density of low-lying discrete levels. Parameters recommended for use in HF calculations are tabulated in
Ref.~\cite{RIPL}. The global parameter systematics for both the Fermi-gas and Gilbert-Cameron formulas have been
developed in Ref.~\cite{Egidy}.  However, it is still unclear how well these parameters reproduce compound
reaction cross sections. No systematic investigations have been performed yet. Experimental data on level
density above discrete levels are scarce. Some information is available from particle evaporation spectra
(i.e. from compound nuclear reactions).
The latest data obtained from (p,n) reaction on Sn isotopes~\cite{Zhur} claims that the available
level density parameters do not reproduce neutron cross sections thereby indicating the problem with level density
parameterizations. It becomes obvious that more experimental data on level density are needed in the energy region
above discrete levels.

In this work, we study compound nuclear reactions to obtain information about level densities of the residual nuclei from
particle evaporation spectra. Two different reactions, d+$^{59}$Co and $^3$He+$^{58}$Fe, which produce the same
$^{61}$Ni compound nucleus, have been investigated. This approach helps to eliminate uncertainties connected to a specific
reaction mechanism. As opposed to most of the similar experiments where only one type of outgoing particles has
been measured, we have measured cross sections of all main outgoing particles including neutrons, protons, and
$\alpha$-particles, populating $^{60}$Ni, $^{60}$Co, and $^{57}$Fe, respectively.

We will begin with a discussion of the present status of level density
estimates used as inputs for HF calculations.

\section{Methods of level density estimates for HF codes}

The simple level counting method to determine the level density of a nucleus works only up to a certain excitation
energy below which levels are well separated and can be determined from nuclear spectroscopy. This region is
typically up to 2~MeV for heavy nuclei and up to 6-9 MeV for light ones. Above these energies, more sophisticated
methods have to be applied.

\subsection{Level density based on neutron resonance spacings}

In the region of neutron resonances which are located just above the neutron binding energy ($B_n$), the level
density can again be determined by counting. In this case neutron resonances are counted; one must also take into
account the assumed spin cut-off factor $\sigma$. Traditionally, because of the the absence of reliable data
below B$_n$, the level density is determined by an interpolation procedure between densities of low-lying discrete
levels and density obtained from neutron resonance spacing.The Bethe Fermi-gas model~\cite{Bethe}
with adjustable parameters $a$ and $\delta$ is often used as an interpolation formula:
\begin{equation}
 \rho(E)=\frac{\exp[2\sqrt{a(E-\delta)}]}{12\sqrt{2}\sigma a^{1/4}(E-\delta)^{5/4}},
\label{eq:rho}
\end{equation}
where the $\sigma$ is the spin cut-off factor determining the level spin distribution. There are a few drawbacks
to this approach. One shortcoming is that it uses an assumption that the selected model is valid in the
entire excitation energy region including low-lying discrete states and neutron resonances. Undoubtedly this is
correct for some of the nuclei. A nice example is the level density of $^{26}$Al which exhibits Fermi-gas
behavior up to 8~MeV of excitation energy~\cite{Egidy}. On the other hand the level densities of $^{56,57}$Fe
measured with Oslo method~\cite{femo}, for example, show complicated behavior  which cannot be described by simple
Fermi-gas formula. The reason for this might be the influence of pairing correlations leading to step structures in
vicinity of proton and neutron paring energies and above. In such cases the model function fit to discrete levels
may undergo considerable deviations in the higher excitation energy region leading to incorrect determination of
level density parameters.

Another consideration is associated with the spin cut-off parameter which is important in determination of the
total level density from density of neutron resonances at $B_n$. In Fermi-gas model the spin cut-off parameter is
determined according to:
\begin{equation}
\sigma^2=\overline{m^2}gt=\frac{I}{\hbar^2}t,
\end{equation}
where $\overline{m^2}$ is the average of the square of the single particle spin projections,
$t=\sqrt{(E-\delta)/a}$ is the temperature, $g=6a/\pi^2$ is the single particle level density, $I$ is the rigid
body moment of inertia expressed as $I=(2/5)\mu AR^2$, where $\mu$ is the nucleon mass, $A$ is the mass number and
$R=1.25A^{1/3}$ is the nuclear radius. The spin cut off parameter in rigid body model is :
\begin{equation}
\sigma_1^2=0.0146A^{5/3}t=0.0146A^{5/3}\sqrt{((E-\delta)/a))}. \label{eq:rigid}
\end{equation}

On the other hand the Gilbert and Cameron \cite{GC} used $\overline{m^2}=0.146A^{2/3}$. The corresponding formula
for $\sigma$ is:

\begin{equation}
\sigma_2^2=0.089A^{2/3}a\sqrt{((E-\delta)/a)}. \label{eq:half}
\end{equation}
Eqs.~(\ref{eq:rigid}) and~(\ref{eq:half}) have the same energy and $A$ dependence ($\sigma^2 \sim
A^{7/6}(E-\delta)^{1/2}$) but differ by a factor of $\approx 2$. It should be mentioned also that the recent model
calculations \cite{alhassid} show the suppression of the moment of inertia at low temperatures compared to its
rigid body value. Thus uncertainties in spin cut off parameter transform to corresponding uncertainties of total
level densities derived from neutron resonance spacings.

Experimentally, the spin cutoff parameter can be obtained only from spin distribution of low-lying discrete
levels. However, because of the small number of known spins, the uncertainty of such procedure is large. It turns
out that reported systematics based on such investigation $\sigma=(0.98\pm0.23)A^{(0.29\pm0.06)}$ \cite{Egidy} is
different from above expressions for which $\sigma \sim~\sqrt{A^{7/6}}~=~A^{0.58}$. At higher excitation energies
determining the cutoff parameter becomes problematic due to the high level density and the absence of the reliable
observables sensitive to this parameter. One can mention Ref.~\cite{grimes} where the spin cutoff parameter has been
determined from the angular distribution of evaporation neutrons with $\alpha$ and proton projectiles. The
deviation from the expected $A$ dependence has also been reported.  The absolute values of the parameter agree with
Eq.~(\ref{eq:rigid}).

The parity dependence of level densities is also not established experimentally beyond the discrete level
region. At the neutron binding energy the assumption is usually made about the equality of negative and positive
parity states. This is supported by some experimental results~\cite{Mitchell}. However, recent calculations,
performed for Fe, Ni, and Zn isotopes, show that for some of them the assumption of equally distributed states is
not fulfilled even far beyond the neutron binding energy, up to excitation energies 15-20 MeV~\cite{rouser2}.

As is seen from the above considerations, the calculation of the total level density from neutron resonance spacing
might contain uncertainties associated with many factors such as the possible deviation from Fermi-gas dependence
in interpolation region, uncertainties in spin cutoff parameter and inequality of states with different parity.
Thus the question of how large these uncertainties are or to what extent the level density extracted in such a way
can be applicable to calculations of reaction cross sections still remains important and not completely resolved.

\subsection{Level density from evaporation particles}

The cross section of evaporated particles from the first stage of a compound-nuclear reaction (i.e. when the
outgoing particle is the first particle resulting from compound nucleus decay ) can be calculated in the framework
of the Hauser-Feshbach theory:
\begin{eqnarray}
\lefteqn{\frac{d\sigma}{d\varepsilon_b}(\varepsilon_a,\varepsilon_b)=}\\ \label{eq:HF}
&&\sum_{J\pi}\sigma^{\mathrm{CN}}(\varepsilon_a)\,\frac{\sum_{I\pi}
\Gamma_b(U,J,\pi,E,I,\pi)\rho_b(E,I,\pi)}{\Gamma(U,J,\pi)}\nonumber
\end{eqnarray}
with
\begin{eqnarray}
\lefteqn{\Gamma(U,J,\pi)=\sum_{b^\prime}\left(\sum_k \Gamma_{b^\prime}(U,J,\pi,E_k,I_k,\pi_k)\,+\right.}\\
&&\left.\sum_{I^\prime\pi^\prime}\int_{E_c}^{U-B_{b^\prime}}dE^\prime\,
\Gamma_{b^\prime}(U,J,\pi,E^\prime,I^\prime,\pi^\prime)\,
\rho_{b^\prime}(E^\prime,I^\prime,\pi^\prime)\right).\nonumber
\end{eqnarray}

Here $\sigma^{CN}(\varepsilon_a)$ is the fusion cross section, $\varepsilon_a$ and $\varepsilon_b$ are energies of
relative motion for incoming and outgoing channels ($\varepsilon_b=U-E_k-B_b$, where $B_b$ is the separation
energy of particle $b$ from the compound nucleus), the $\Gamma_{b}$ are the transmission coefficients of the
outgoing particle, and the quantities $(U,J,\pi)$ and $(E,I,\pi)$ are the energy, angular momentum, and parity of
the compound and residual nuclei, respectively. The energy $E_c$ is the continuum edge, above which levels are
modeled using a level density parameterization. For energies below $E_c$ the known excitation energies, spins,
and parities of discrete levels are used. In practice $E_c$ is determined by the available spectroscopic data in
the literature. It follows from Eq.~(\ref{eq:HF}) that the cross section is determined by both transmission
coefficients of outgoing particles and the NLD of the residual nucleus $\rho_b(E,I,\pi)$\@. It is believed that
transmission coefficients are known with sufficient accuracy near the line of stability because they can be
obtained from optical model potentials usually based on experimental data for elastic scattering and total cross
sections in the corresponding outgoing channel. Transmission coefficients obtained from different systematics of
optical model parameters do not differ by more that 15-20 \% from each other in our region of interest ($1-15$~MeV
of outgoing particles). The uncertainties in level densities are much larger. Therefore the Hauser Feshbach model
can be used to improve level densities by comparing experimental and calculated particle evaporation spectra.
Details and assumptions of this procedure are described in Refs~\cite{Vonach,Vonach1}.

The advantage of this method is that because of the wide range of spin population in both the compound and final
nuclei, evaporation spectra are determined by the total level density (integrated over all level spins) as opposed
to the neutron resonance technique where resonances are known for one or two spins and one parity. The drawback
stems from possible direct or multistep compound reaction contributions distorting the evaporation spectra,
especially in the region of low-lying discrete levels needed for the absolute normalization of obtained level
densities.

 According to Hodgson \cite{Hudson}, the interaction process can usefully be considered to take place in a series
of stages corresponding to the successive nucleon-nucleon interaction until complete equilibrium is reached. At
each stage it is possible for particles to be emitted from the nucleus. The direct reactions refer to the fast,
first stages of this process giving forward peaked angular distribution. The term multistep direct reaction
implies that that such process may take place in a number of states. Compound nuclear reactions refers to all
processes giving angular distributions symmetric about 90$^0$; they are subdivided into multistep compound
reactions that take place before the compound system has attained final statistical equilibrium and statistical
compound reactions that correspond to the evaporation of particle from an equilibrium system.

The use of evaporation spectra to infer level densities requires that the reaction goes through to complete
equilibrium. Significant contributions from either multistep direct or multistep compound reactions could cause
incorrect level density parameters to be deduced.  Multistep direct reactions would usually be forward peaked and
also concentrated in peaks.  If the reaction has a limited number of stages, the two-body force cannot cause
transitions to states which involve a large number of rearrangements from the original state. Multistep compound
reactions would be expected to lead to angular distributions which are symmetric about 90$^0$.  They would, if
complete equilibration has not occurred, also preferentially reach states which are similar to the target plus
projectile. The shape of spectra from a multistep compound reaction would be different for a deuteron-induced as
opposed to a $^3$He-induced reaction. Hodgson has reviewed \cite{Hudson} the evidence for multistep compound
reactions. He finds the most convincing evidence for such contributions comes from fluctuation measurements for
the $^{27}$Al($^3$He,p) reaction.  In this case, certain low-lying states show level widths in the compound system
which are larger than expected.  These states are low-lying and are the ones which would be most likely to show
such effects.  It appears that measurements of continuum spectra do not show evidence of such contributions. The
uncertainties connected to contributions of pre-equilibrium reactions are generally difficult to estimate
experimentally. The measurement of angular distribution does not solve the problem in the case of multistep
compound mechanism. We believe that the use of different reactions to form the same compound nucleus is the most
reliable way to estimate and eliminate such contributions.

In this work we investigate reactions with deuteron and $^3$He projectiles on $^{59}$Co and $^{58}$Fe,
respectively.  These two reactions form the same compound nucleus, $^{61}$Ni. The purpose was to investigate if the
cross section of outgoing particles from both reactions can be described in the framework of Hauser-Feshbach model
with same set of level density parameters. This is possible only when production cross section is due to compound
reaction mechanism in both reactions. Neutron, protons, and $\alpha$-particles have been measured. These outgoing
particles exhaust the majority of the fusion cross section. The ratio between cross sections of different
particles is determined by the ratio of level densities of corresponding residual nuclei. It puts constraints on
relative level density values obtained from an experiment. In our experiment, the level densities of $^{60}$Ni,
$^{60}$Co, and $^{57}$Fe residual nuclei have been determined from the region of the energy spectra of neutron,
proton, and $\alpha$-particles where only first state emission is possible.

\section{Experiment and method}

\begin{figure}[htb!]
\vspace{1.5cm}
\includegraphics[width=7.0cm]{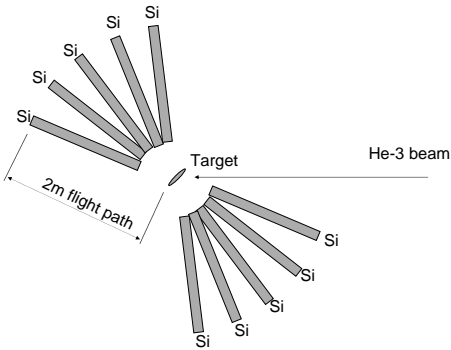}
\vspace{-1.5cm} \caption{Charge particle spectrometer utilized for the measurements.}\label{fig:sp}
\end{figure}

The tandem accelerator at Ohio University's Edwards Accelerator Laboratory provided
$^3$He and deuteron beams with energies of 10 and 7.5~MeV, respectively. Self-supporting foils of 0.625-mg/cm$^2$
$^{58}$Fe (82\% enriched) and 0.89-mg/cm$^2$ ${}^{59}{\rm Co}$ (100\% natural abundance) have been used as
targets. The outgoing charged particles were registered by charged-particle spectrometers as shown in
Fig.~\ref{fig:fig1}. The setup has ten 2-m time-of-flight legs ending with Si detectors (see Fig.~\ref{fig:sp}). Legs are
set up at different angles ranging from $22.5^\circ$ up to $157.5^\circ$. The mass of the charged particles is
determined by measuring both the energy deposited in Si detectors and the time of flight. Additionally, a neutron
detector was placed at the distance of 140~cm from the target to measure the neutron energy spectrum. The mass
resolution was sufficient to resolve protons, deuterons, ${}^3{\rm H}$/$^3$He, and $\alpha$-particles.

Additionally, the neutron spectra from both the $^{58}$Fe$(^3$He$,Xn)$ and the $^{59}$Co$(d,Xn)$ reactions have
been measured by the time-of-flight method with the Swinger facility of Edwards Laboratory~\cite{Swinger}. Here a
flight path of $7$~m has been used to obtain better energy resolution for outgoing neutrons, allowing us to measure
the shape of neutron evaporation spectrum more accurately. The energy of the outgoing neutrons is determined by
time-of-flight method. The 3-ns pulse width provided an energy resolution of about 100~keV and 800~keV at 1 and
14~MeV of neutrons, respectively. The neutron detector efficiency was measured with neutrons from the
$^{27}$Al$(d,n)$ reaction on a stopping Al target at $E_d=7.44$~MeV~\cite{Aleff}. This measurement allowed us to
determine the detector efficiency from 0.2 to 14.5~MeV neutron energy with an accuracy of $\sim 6$\%. The neutron
spectra have been measured at backward angles from $110^\circ$ to $150^\circ$. Additional measurements with a
blank target have been performed at each angle to determine background contribution. The absolute cross section
has been calculated by taking into account the target thickness, the accumulated charge of incoming deuteron or
$^3$He beam, and the neutron detector efficiency. The overall systematic error for the absolute cross sections
is estimated to be 15\%. The errors in ratios of proton and $\alpha$ cross sections are only a few percents because
they are determined by counting statistics alone.

\section{Experimental particle spectra and level density of product nuclei}

\begin{figure*}
\includegraphics[width=16.5cm]{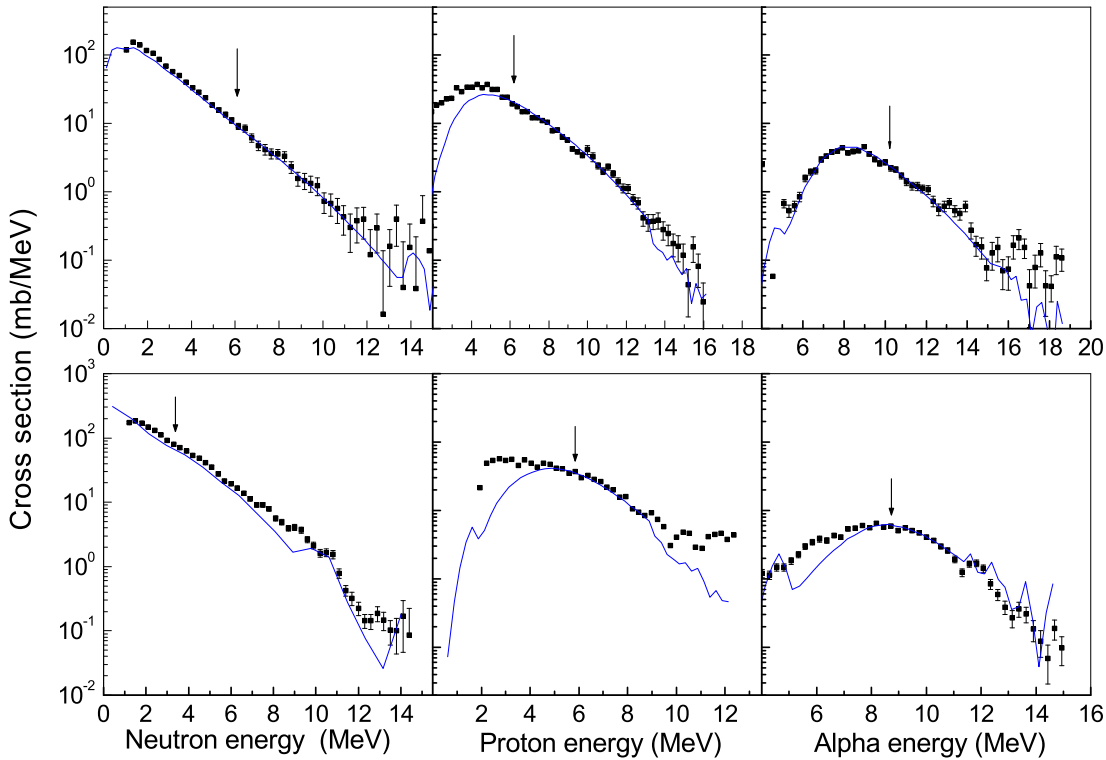}
\caption{Particle energy spectra for the  $^3$He+$^{58}$Fe (upper panel)and d+$^{59}$Co (lower panel) reactions.
The experimental data are shown by points. Solid lines are HF calculations with level density parameters extracted
from the experiment. Calculations have been multiplied by reduction factor $K=0.52$ due to direct reaction
contributions. Arrows show energies above which spectra contain only contributions from the first stage of the
reaction.} \label{fig:fig1}
\end{figure*}

Energy spectra of neutron, protons, and $\alpha$-particles have been measured at backward angles (from $112^\circ$
to $157^\circ$) to eliminate contributions from direct reaction mechanisms. Fig.~\ref{fig:fig1} show energy
spectra of outgoing particles for both the $^3$He+$^{58}$Fe and $d$+$^{59}$Co reactions. The calculations of
particle energy spectra have been performed with Hauser-Feshbach (HF) program developed at Edwards Accelerator Lab
of Ohio University \cite{INPP}. Particle transmission coefficients have been calculated with optical model
potentials taken from the RIPL data base \cite{RIPL}. Different potentials have been tested and found to be the
same within 15\%. Alpha-particle potentials are more uncertain. Differences between corresponding
$\alpha$-transmission coefficients depends on the $\alpha$-energy and varies from $\sim~40$\% for lower
$\alpha$-energies to $<1$\% for higher $\alpha$-energies in our region of interest (8-18~MeV). In order to reduce
these uncertainties the RIPL $\alpha$-potentials have been tested against the experimental data on low energy
$\alpha-$elastic scattering on $^{58}$Ni \cite{alphael}. The data have been reproduced best by the potential from
Ref.\cite{9019} which has been adopted for our HF calculations. Four level density models have been chosen for
testing:
\begin{itemize}
\item The M1 model uses the Bethe formula (\ref{eq:rho}) with parameters adjusted to fit both discrete level
density and neutrons s-wave resonance spacings.
\item The M2 model uses the Gilbert-Cameron \cite{GC} formula with parameters adjusted to fit both discrete level
density and neutrons s-wave resonance spacings.
\item The M3 model uses Bethe formula but $\delta$ parameters are obtained from
pairing energies according to Ref.~\cite{rauscher}. The $a$ parameter has been adjusted to match s-wave neutron
resonance spacing. This model does not fit discrete levels.
\item The M4 model is based on microscopic Hartree-Fock-BCS calculations~\cite{HFBCS} which are available from RIPL
data base~\cite{RIPL}. According to Ref.~\cite{HFBCS}, this model has also been renormalized to fit discrete
levels and neutron resonance spacings.
\end{itemize}
The value of total level density derived from neutron resonance spacings depends on spin cutoff parameter used.
Therefore two prescriptions (\ref{eq:rigid}) and (\ref{eq:half}) for this parameter have been tested for M1-M3
models. The M4 model uses its one spin distribution which it is close to the prescription (\ref{eq:rigid}).

The measured particle energy spectra include particles from all possible stages of the reaction. However, by
limiting our consideration to particles with energies above a particular threshold, we can ensure that only
particles from the first stage of the reaction contribute. These thresholds depend on the particular reaction and
are indicated by the arrows in Fig.~\ref{fig:fig1}. In this energy interval cross sections are determined
exclusively by the level density of those residual nuclei. Another aspect which should be taken into consideration
when comparing calculations and experiment is the contribution of direct processes. Direct processes take away the
incoming flux resulting in reduction of compound reaction contribution. Assuming that the total reaction cross
section ($\sigma_R$) can be decomposed into the sum of direct ($\sigma_r^d$) and compound reaction mechanisms
($\sigma_R^c$), we have $\sigma_R=\sigma_R^{d}+\sigma_R^{c}$. In this case, the HF calculations should be
multiplied by the constant factor $K=\sigma_R^{c\_exp}/\sigma_R$ to correct for the absorbed incident flux which
does not lead to compound nucleus formation. In our experiment the K has been estimated from the ratio
$K^{exp}=(\sigma_n^{exp}+\sigma_p^{exp}+\sigma_\alpha^{exp})/(\sigma_n^{calc}+\sigma_p^{calc}+\sigma_\alpha^{calc})
\approx K$ where the experimental cross sections have been measured at backward angles. If level densities used in
calculations are correct, $K^{exp}=K$. However, the calculations show that this parameter is not very sensitive to
input level densities and can be estimated with $\sim$20\% accuracy with any reasonable level density models.

Table~\ref{tab:tab1} shows the ratio of theoretical and experimental cross sections for different level
density models used in calculations. Calculations have been multiplied by reduction factor K which for both
reactions varied within 0.48-0.54 for different level density models. Results show that all of the models
reproduce neutron cross sections within $\sim$20\%. However, they overestimate $\alpha$-particle cross sections by
$\sim$30\% in average and underestimate protons by 5-80\%. None of the models reproduce the ratio of p/$\alpha$
cross section; for example all models systematically overestimate this ratio by a factor of $\sim$2 for the d+$^{59}$Co reaction.
Assuming that particle transmission coefficients are known with sufficient accuracy, we conclude that the level
density of residual nuclei is responsible for such disagreement. In particular, the level density ratio
$\rho[{^{57}Fe}]/\rho[{^{60}Co}]$ is overestimated by model calculations.

\begin{table*}[htb!]
\caption {Ratio of experimental  and calculated cross sections obtained with four prior level density models M1-M4
and one posterior M1$^{exp}$ which uses parameters fit to experimental level densities (see Table \ref{tab:tab2}).
The spin cutoff parameters $\sigma_1$ and $\sigma_2$ are defined according to Eqs.~(\ref{eq:rigid}) and~(\ref{eq:half}).}
\begin{tabular}{|c|ccccccc|cc|} \hline
    &\multicolumn{2}{c}{M1} &\multicolumn{2}{c}{M2} &\multicolumn{2}{c}{M3} &\multicolumn{1}{c|}{M4} &M1$^{exp}$ &K$^{exp}$   \\
 &$\sigma_1$ &$\sigma_2$ &$\sigma_1$ &$\sigma_2$ &$\sigma_1$ &$\sigma_2$ &             &  &\multicolumn{1}{c|}{}    \\ \hline
 $^{58}$Fe($^3$He,n)        &0.79(12) &1.04(16) &1.03(16) &1.22(19) &1.03(16) &1.05(16) &0.90(14)   &1.03(16) &0.52  \\
 $^{58}$Fe($^3$He,p)        &1.23(20) &1.01(15) &1.05(16) &0.93(14) &1.03(15) &1.01(15) &1.11(17)   &0.98(15) & \\
 $^{58}$Fe($^3$He,$\alpha$) &0.66(10) &0.81(12) &0.66(10) &0.86(13) &0.73(11) &0.81(12) &0.72(11)   &1.01(15) &  \\ \hline
  $^{59}$Co(d,n)            &0.81(12) &0.90(14) &0.84(13) &0.91(14) &0.92(14) &0.93(14) &0.89(13)   &0.97(15) &0.53    \\
  $^{59}$Co(d,p)            &1.82(27) &1.42(21) &1.70(26) &1.40(21) &1.32(20) &1.24(19) &1.41(21)   &1.07(16) & \\
  $^{59}$Co(d,$\alpha$)     &0.69(11) &0.59(10) &0.64(10) &0.57(10) &0.59(9)  &0.70(11) &0.64(10)   &0.97(15) &  \\ \hline
\end{tabular}
\label{tab:tab1}
\end{table*}

In order to obtain correct level densities, the following procedure has been used as described in
Ref.~\cite{Mndn}. The NLD model is chosen to calculate the differential cross section of Eq.~(\ref{eq:HF}). The
parameters of the model were adjusted to reproduce the experimental spectra as closely as possible. The input NLD
was improved by binwise renormalization according to the expression:
\begin{equation}
\rho_b(E,I,\pi)=\rho_b(E,I,\pi)_{\mathrm{input}} \frac{(d\sigma/d\varepsilon_b)_{\mathrm{meas}}}
{(d\sigma/d\varepsilon_b)_{\mathrm{calc}}}. \label{eq:ld}
\end{equation}
The absolute normalization of the improved level densities (later referred to as experimental level densities)
has been obtained by using
discrete level densities of $^{60}$Ni populated by neutrons from the $^{59}$Co(d,n) reaction. Protons and
$\alpha$-particles populating discrete levels behave differently for different reactions. The Fig. \ref{fig:fig1}
shows that the ratio between experiment and calculations in discrete energy region is greater for $^{59}$Co(d,p)
compared to $^{58}$Fe($^3$He,p) and for $^{58}$Fe($^3$He,$\alpha$) compared to $^{59}$Co(d,$\alpha$). These
enhancements are apparently reaction specific and connected to contribution of direct or/and multistep compound
reaction mechanism. We are not able to make the same comparison for neutron spectra because the counting
statistics in the region of discrete levels for $^{58}$Fe($^3$He,n) reaction are rather poor. However, our recent
result from $^{55}$Mn(d,n) \cite{Mndn} indicates that the neutron spectrum measured at backward angles is purely
evaporated even for high energy neutrons populating discrete levels. Therefore we used the neutron spectrum from
the $^{59}$Co(d,n) reaction to determine the absolute normalization of the level density for the residual nucleus
$^{60}$Ni. The absolute level densities of both $^{60}$Co and $^{57}$Fe nuclei have been adjusted in such a way as
to reproduce ratios of both neutron/proton and neutron/alpha cross sections. Uncertainties of obtained level
densities have been estimated to be about 20\% which include uncertainties of absolute cross section measurements
and uncertainties of particle transmission coefficients.

\begin{figure*}
\includegraphics[width=18.5cm]{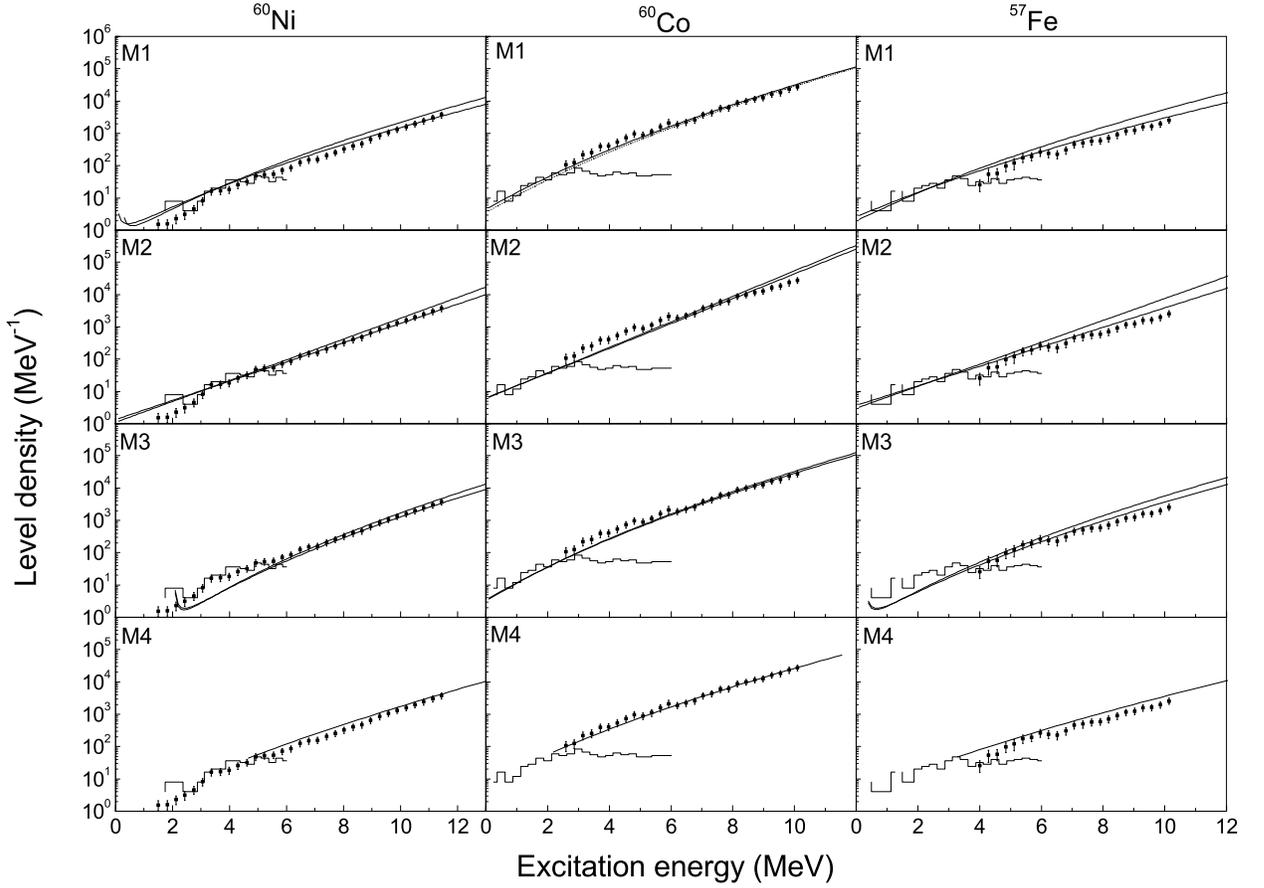}
\caption{Our experimental level density are shown as points. Curves indicate level densities from the four model prescriptions
M1-M4. The upper and lower curves for M1-M3 relate to two spin cutoff parameters $\sigma_1$ and $\sigma_2$ used to determine
total level densities from neutron resonance spacings. The histogram is the density of discrete levels.} \label{fig:fig2}
\end{figure*}

Both experimental and calculated level densities are displayed in Fig.~\ref{fig:fig2}. The level density for
$^{60}$Ni has been extracted from (d,n) spectra because of better counting statistics but ($^3$He,p) and
($^3$He,$\alpha$) reactions have been used to obtain level density for $^{60}$Co and $^{57}$Fe, respectively,
because of larger $Q$ value. This approach allows one to obtain level densities in a larger excitation energy
interval. Calculations have been performed with models M1-M4 with spin cutoff parameters $\sigma_1$ and $\sigma_2$
for M1-M3 models. The M4 model uses its own spin distribution which is close to $\sigma_1$ for these nuclei. The
$\chi^2$ values for calculated and experimental level densities are shown in the Table \ref{tab:hi2}. Results show
that the M1 model with $\sigma_1$ gives worse agreement with experimental data. The use of $\sigma_2$ improve the
agreement for all of the models. The M2 and M3 models using $\sigma_2$ give best agreement with experiment on
average, however level density for $^{60}$Co agrees better when using $\sigma_1$ and the best agreement is reached
with M4 model. It appears that the spin cutoff parameter is very important when deriving the total level density
from neutron resonance spacings. However, none of the models give a perfect description of the experimental data.

In order to improve level density parameters, the experimental level densities have been fitted with the Fermi-gas
function (\ref{eq:rho}) for two different spin cutoff factors $\sigma_1$ and $\sigma_2$. Best fit parameters are
presented in the Table \ref{tab:tab2}. They allow one to reproduce both shapes of particle spectra
(fig.\ref{fig:fig1}) as well as ratios of neutron, proton and $\alpha$ cross sections for both $^3$He+$^{58}$Fe
and $d$+$^{59}$Co reactions (Table \ref{tab:tab1}). Level density parameters have been adjusted independently for
both spin cutoff parameters resulting in the approximately same final ratio of experimental/calculated cross
sections and $\chi^2$ values. Therefore the only one entry M1$^{exp}$ is presented in tables. The fact that a
single set of level density parameters allows one to reproduce all particle cross sections from both reactions
supports our conclusion that the compound nuclear mechanism is dominant in these reactions. Finally we note that
the HF calculations do not perfectly reproduce the low-energy regions of the proton spectra where the second stage
of outgoing protons dominate. Here the calculations also depends on additional level densities of corresponding
residual nuclei as well as on the $\gamma$-strength functions. We leave this problem for further investigations.

\begin{table}[htb!]
\caption {Fermi-gas parameters obtained from experimental level densities}
\begin{tabular}{|c|c|c|c|} \hline
 Nucleus  & $^{60}$Ni & $^{60}$Co &$^{57}$Fe \\ \hline
$a,\delta$ for Eq.(\ref{eq:rigid}) &6.16;1.43 &6.91;-1.89 &5.92;-0.13 \\ \hline $a,\delta$ for Eq.(\ref{eq:half})
&6.39;0.80 &7.17;-2.6 &6.14;-0.78 \\ \hline
\end{tabular}
\label{tab:tab2}
\end{table}

\begin{figure}[htb!]
\includegraphics[width=8.5cm]{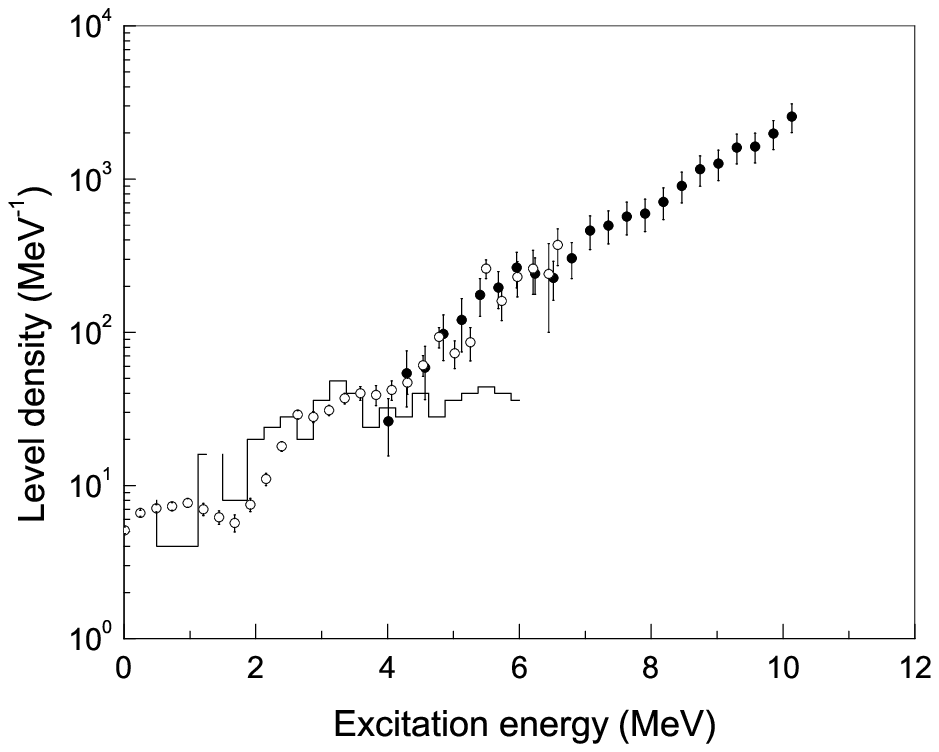}
\caption{The experimental level densities of $^{57}$Fe nucleus. Filled points are present experimental values.
Open points are data from Oslo experiment~\cite{femo}. Histogram is density of discrete levels.} \label{fig:fig3}
\end{figure}

The level density of $^{57}$Fe below the particle separation threshold has also been obtained \cite{femo} by Oslo
technique using particle-$\gamma$ coincidences from $^{57}$($^3$He,$^3$He$^\prime$)$^{57}$Fe reaction. We performed
a similar comparison for the $^{56}$Fe nucleus where we confirmed consistency of both the Oslo technique and the
technique based on particle evaporation spectra. Figure~\ref{fig:fig3} shows the comparison for the $^{57}$Fe
nucleus. Here we also see good agreement between level densities obtained from two different experiments. It
supports the obtained level densities.
  The Fermi-gas parameters for $^{60}$Ni have been obtained in Ref.~\cite{ni60} from
  $^{63}$Cu(p,$\alpha$)$^{60}$Ni reaction at E$_p$=12~MeV . The values a=6.4 and $\delta$=1.3 are in a good
  agreement with our parameters presented in the Table \ref{tab:tab2}.
\begin{table} [htb!]
\caption{$\chi2$ of experimental and calculated total level densities for different level density models and spin
cutoff factors}
\begin{tabular}{|c|ccccccc|c|} \hline
 Nucleus &\multicolumn{2}{c}{M1} &\multicolumn{2}{c}{M2} &\multicolumn{2}{c}{M3} &M4 &M1$^{exp}$ \\
         &$\sigma_1$ &$\sigma_2$ &$\sigma_1$ &$\sigma_2$ &$\sigma_1$ &$\sigma_2$ & &    \\ \hline
 $^{60}$Ni &15.3 &3.8  &1.5  &0.2 &0.9  &1.4 &2.5 &0.6 \\
 $^{60}$Co &1.3  &1.9  &2.9  &3.1 &1.8  &2.0 &0.8 &0.6 \\
 $^{57}$Fe &20.2 &2.5  &18.9 &2.0 &10.8 &1.8 &5.8 &0.6 \\ \hline
All nuclei &11.5 &2.7  &7.5  &1.8 &4.3  &1.8 &3.1 &0.6 \\ \hline
\end{tabular}
\label{tab:hi2}
\end{table}

\section{Spin cutoff parameter}

As it has been mentioned in the previous section the spin cutoff parameter $\sigma_2$ obtained according to
Eq.~(\ref{eq:half}) gives slightly better agreement with the experiment compared to $\sigma_1$ obtained from
Eq.~(\ref{eq:rigid}). On the other hand, the spin cut off parameters at the neutron binding energy can be directly
obtained from the experimental total level density and the density  of levels for one or several spin states which
are known from the analysis of neutron resonances~\cite{RIPL}. We used the spin distribution formula from Ref.~\cite{GC}:
\begin{equation}
G(J)= \frac{(2J+1)}{2\sigma^2}\exp{\Bigl[\frac{-(J+0.5)^2}{2\sigma^2}\Bigr]}\label{eq:spin}
\end{equation}
with normalization condition:
\begin{equation}
\sum_J{G(J)}=1
\end{equation}
The total level density $\rho(U)$ can be connected to the neutron resonance spacing by using the expression:
\begin{equation}
\frac{2}{D_L}=\rho(B_n+0.5\Delta E)\sum_{J=|I_0-0.5-L|}^{|I_0+0.5+L|}{G(J)}, \label{eq:spin1}
\end{equation}
where $D_L$ is the neutron resonance spacing for neutrons with orbital momentum $L$, $\Delta E$ is the energy
interval containing neutron resonances. The assumption of equality of level numbers with positive and negative
parity is used. Because the total level density $\rho(B_n+0.5\Delta E)$ around the neutron separation energy is
known from our experiment, the parameter $\sigma$ can be obtained from Eqs.(\ref{eq:spin})-(\ref{eq:spin1}).

The data on neutron resonance spacings for nuclei under study  are taken from Ref.~\cite{Atlas}. The estimated spin
cut off parameters from both $s$-wave ($L=0$) and $p$-wave ($L=1$) resonance spacings are presented in the
Table~\ref{tab:tab4}. The uncertainties include a 20\% normalization uncertainty in total level densities and
uncertainties in the resonance spacings. For $^{57}$Fe, we have obtained good agreement between two values of $\sigma$
derived from s- and p-wave neutron resonances. It indicates the parity equilibrium of neutron resonances. For
$^{60}$Co, the uncertainties are too large to draw a definite conclusion. For $^{60}$Ni, the only $D_0$ is known
and one value of $\sigma$ is obtained. It agrees better with $\sigma_2$ but $\sigma_1$ cannot be excluded.

The calculations of spin cutoff parameter have been performed with Eqs.~(\ref{eq:rigid}) and~(\ref{eq:half})
with Fermi-gas parameters from Table \ref{tab:tab2}. The experiment shows better agreement with $\sigma_2$ for
$^{57}$Fe. Spin cut off parameters for $^{60}$Ni and $^{60}$Co agree better with $\sigma_2$ and $\sigma_1$
respectively, however because of the large uncertainties, it is impossible to draw an unambiguous conclusion.

\begin{table}[htb!]
\caption{Spin cut off parameter obtained from $s$-wave $\sigma^{exp}_s$ and p-wave $\sigma^{exp}_p$ resonances
with using the total level density from the experiment. $\sigma^{cal}_1$ and $\sigma^{cal}_2$ have been calculated
according to Eqs.~(\ref{eq:rigid}) and~(\ref{eq:half}), respectively, with parameters from
Table~\ref{tab:tab2}.}
\begin{tabular}{|c|ccc|} \hline
Nucleus &$^{60}$Ni &$^{60}$Co  &$^{57}$Fe  \\ \hline
 $\sigma^{exp}_s$ &3.3(8)  &3.6(15) &2.80(30) \\
 $\sigma^{exp}_p$ &        &5.2(12) &2.88(35) \\
 $\sigma^{cal}_1$ &4.13     &3.95     &3.76   \\
 $\sigma^{cal}_2$ &3.22     &3.26     &3.0  \\ \hline
\end{tabular}
\label{tab:tab4}
\end{table}

\section{Discussion}
The consistency between results from two different reactions supports our conclusion that these reactions are
dominated by the compound-nuclear reaction mechanism at backward angles.
Our results show that level densities estimated on the basis of
interpolation procedure between neutron resonance and discrete energy regions do not reproduce experimental cross
sections of all outgoing particles simultaneously. The reason is that for some of the nuclei the level density
between discrete and continuum regions has a complicated behavior which cannot be described by simple formulas
based on Fermi-gas or Gilbert-Cameron models. It is seen for $^{57}$Fe (Fig.~\ref{fig:fig3}) where the level
density exhibits some step structure at energy around 3.7~MeV. Nevertheless, the Fermi-gas model can still be used
to describe the level density at higher excitation energies where density fluctuations vanish. The M3 model, which
does not use discrete levels, gives best agreement. However, a problem apparently connected to the spin cutoff
parameters is still present. These results indicate that it is necessary to use level density systematics obtained
from compound-nuclear particle evaporation spectra. Obviously, the region of discrete levels should be excluded
from such an analysis.

Spin cut off parameters obtained from this experiment are in general agreement with model prediction of
Eqs.(\ref{eq:rigid}) and (\ref{eq:half}). However it is difficult to reduce uncertainties to make more specific
conclusions about the origin of this parameter. Most probably, this parameter fluctuates from nucleus to nucleus
and is determined by the internal properties of nuclei such as the specific population of shell orbits.

As it has been discussed in the introduction, level densities affect reaction rates which are important in
astrophysics and other applications. The magnitude of this affect depends mainly on level densities and
contribution of the channel of interest to the total reaction cross section.  According to the Table
\ref{tab:tab1}, the neutron outgoing channel is less sensitive to variations of level densities while changes in
proton and $\alpha$ cross sections can reach a factor of 2 from corresponding changes in level densities. Changes
in predicted cross sections will also occur at this level.

\section{Conclusion}

The neutron, proton, and $\alpha$-particle cross sections have been measured at backward angles from
$^3$He+$^{58}$Fe and d+$^{59}$Co reactions. The calculations using HF model have been performed with three level
density models adjusted to match discrete levels and neutron resonance spacings and one model adjusted to match
neutron resonances only. None of the model reproduces cross sections of all outgoing particles simultaneously from
both reactions. However, the model M3 suggested in Ref.\cite{rauscher} gives the best agreement with experiment.

Level densities of residual nuclei $^{60}$Ni, $^{60}$Co, and $^{57}$Fe have been obtained from particle
evaporation spectra. Experimental level densities have been fit by Fermi-gas function and new level density
parameters have been obtained. The new level densities allow us to reproduce all particle energy spectra from both
reactions that indicate the dominance of compound nuclear mechanism in particle spectra measured at backward
angles. The contribution of compound mechanism to the total cross section is estimated about 50\% for both reactions.

The total level density obtained from particle spectra and neutron resonance spacings have been used to extract
the spin cut off parameter at the neutron separation energies. The extracted parameters agree with
predictions of Eq.~(\ref{eq:half}) for $^{57}$Fe but no definite conclusions can be made for $^{60}$Ni and $^{60}$Co. A
better understanding of parity ratio systematics would help to make this technique more reliable.

\section{acknowledgments}

We are grateful to J.E.~O'Donnell and D.~Carter for computer and electronic support during the experiment,
A.~Adekola, C.~Matei, B.~Oginni and Z.~Heinen for taking shifts, D.C.~Ingram for target thickness calculations done
for us. We also acknowledge financial support from Department of Energy, grant No. DE-FG52-06NA26187/A000.


\begin{thebibliography}{99}

\bibitem{rauscher} T.~Rauscher, F.K.~Thielemann, K.L.~Kratz, Phys.\ Rev.\ C \bf 56\rm, 1613 (1997).
\bibitem{AFC} Report of the Nuclear Physics and Related Computational Science R\&D for
Advanced Fuel Cycle Workshop, Bethesda Maryland, August 10-12,2006,
http://www-fp.mcs.anl.gov/nprcsafc/Report\_FINAL.pdf

\bibitem{houser} W.~Hauser and H.~Feshbach, Phys.\ Rev. \bf 87\rm, 366 (1952).
\bibitem{Bethe} H.A. Bethe, Phys.Rev. 50, 332(1936).
\bibitem{GC} A.~Gilbert and A.G.W.~Cameron, Can.J.Phys. \bf 43\rm, 269 (1965).
\bibitem{RIPL}T. Belgya, O. Bersillon, R. Capote, T. Fukahori, G. Zhigang, S.
Goriely, M. Herman, A.V. Ignatyuk, S. Kailas, A. Koning, P. Oblozhinsky, V. Plujko, and P. Young, Handbook for
calculations of nuclear reaction data: Reference Input Parameter Library. Available online at
http://www-nds.iaea.org/RIPL-2/, IAEA, Vienna, 2005.
\bibitem{Egidy}T. von Egidy and D. Bucurescu, Phys.\ Rev.\ C \bf 72\rm, 044311 (2005); \bf 73\rm, 049901(E) (2006).
\bibitem{Zhur} B.V. Zhuravlev, A.A.Lychagin, and N.N.Titarenko, Physics of Atomic Nuclei, \bf 69\rm, 363(2006).
\bibitem{femo}A. Schiller {\sl et al.}, Phys.\ Rev.\ C \bf 68\rm, 054326 (2003).
\bibitem {alhassid} Y.~Alhassid, G.F.~Bertsch, L.~Fang, and S.~Liu,  Phys.\ Rev.\ C \bf 72\rm, 064326 (2005).
\bibitem{grimes}S.M.~Grimes, J.D.~Anderson, J.W.~McClure, B.A.~Pohl, and C.~Wong, Phys.\ Rev.\ C \bf 10\rm, p.2373
(1974).
\bibitem{Mitchell} S.J.Lokitz, G.E.Mitchell, J.F.Shriner, Jr, Phys.\ Rev.\ C \bf 71\rm, 064315(2005).
\bibitem{rouser2} D Mocelj, T.Rauscher, F-K Thielemann, G Mart\'{i}nez Pinedo, K.Langanke, L.Pacearescu and
A.F\"{a}\ss ler, J.Phys.G:Nucl.Part. Phys. \bf 31\rm, 1927(2005)
\bibitem{Vonach}H. Vonach, Proceedings of the IAEA Advisory Group Meeting on
\bibitem{Vonach1}A. Wallner, B. Strohmaier, and H. Vonach, Phys.\ Rev.\ C \bf 51\rm, 614 (1994).
\bibitem{Hudson} P.E. Hodgson, Rep. Prog. Phys. \bf 50 \rm 1171(1987).
Basic and Applied Problems of Nuclear Level Densities, Upton, NY, 1983, BNL Report No.\ BNL-NCS-51694, 1983, p.\
247.
\bibitem{Swinger} A. Salas-Bacci, S.M. Grimes, T.N. Massey, Y. Parpottas, R.T. Wheeler, J.E. Oldendick,
Phys.\ Rev.\ C \bf 70\rm, 024311 (2004).
\bibitem{Aleff} T.N. Massey, S. Al-Quraishi, C.E. Brient, J.F. Guillemette, S.M. Grimes,
D. Jacobs, J.E. O'Donnell, J. Oldendick and R. Wheeler, Nuclear Science and Engineering 129, 175 (1998).
\bibitem{INPP} S. M. Grimes, Ohio University Report INPP-04-03, 2004 (unpublished).
\bibitem{alphael} L.R.Gasques, L.C.Chamon, D.Pereira, V.Guimar\~{a}es, A.L\'{e}pine-Szily, M.A.G.Alvarez, E.S.Rossi,
Jr., C.P.Silva, B.V.Carlson, J.J.Kolata, L.Lamm, D.Peterson, P.Santi, S.Vincent, P.A.De Young, G.Peasley,
  Phys.\ Rev.\ C \bf 67\rm, 024602 (2003).
\bibitem{9019} R.C.Harper and W.L.Alford, J.Phys.G:Nucl.Phys. 8, 153(1982)
\bibitem{HFBCS} P.  Demetriou and S.  Goriely, Nucl.\ Phys.\ {\bf A695}, 95 (2001).
\bibitem{Mndn} A.V.Voinov, S.M.Grimes, U.Agvaanluvsan, E.Algin, T.Belgya, C.R.Brune, M.Guttormsen, M.J.Hornish,
T.Massey, G.E.Mitchell, J.Rekstad, A.Schiller, S.Siem, Phys.\ Rev.\ C \bf 74\rm, 014314 (2006).
\bibitem{ni60}  Louis C. Vaz, C.C.Lu, and J.R.Huizenga, Phys.\ Rev.\ C \bf 5\rm, p.463 (1972).
\bibitem{Atlas} S.F. Mughabhab, Atlas of Neutron Resonances, Elsevier 5-th ed. 2006.














\end{thebibliography}
\end{document}